# Extreme multiplex spectroscopy at wide-field 4-m telescopes

Robert Content* & Tom Shanks
Department of Physics, Durham University, South Road, Durham, DH1 3LE, UK

## ABSTRACT

We describe the design and science case for a spectrograph for the prime focus of classical 4-m wide-field telescopes that can deliver at least 4000 MOS slits over a 1° field. This extreme multiplex capability means that 25000 galaxy redshifts can be measured in a single night, opening up the possibilities for large galaxy redshift surveys out to z~0.7 and beyond for the purpose of measuring the Baryon Acoustic Oscillation (BAO) scale and for many other science goals. The design features four cloned spectrographs and exploits the exclusive possibility of tiling the focal plane of wide-field 4-m telescopes with CCDs for multi-object spectroscopic purposes. In ~200 night projects, such spectrographs have the potential to make galaxy redshift surveys of ~$6x10^6$ galaxies over a wide redshift range and thus may provide a low-cost alternative to other survey routes such as WFMOS and SKA. Two of these extreme multiplex spectrographs are currently being designed for the AAT (NG1dF) and Calar Alto (XMS) 4-m class telescopes. NG2dF, a larger version for the AAT 2° field, would have 12 clones and at least 12000 slits. The clones use a transparent design including a grism in which all optics are smaller than the clone square subfield so that the clones can be tightly packed with little gaps between the contiguous fields. Only low cost glasses are used; the variations in chromatic aberrations between bands are compensated by changing one or two of the lenses adjacent to the grism. The total weight and length is smaller with a few clones than a unique spectrograph which makes it feasible to place the spectrograph at the prime focus.

## 1. INTRODUCTION

Recently there has been increasing appreciation of the contribution that small aperture telescopes with wide fields of view can make to astronomical imaging surveys. The Pan-STARRS project is one example. However, there has been less appreciation of smaller telescopes for wide-field multi-object spectroscopy. But the idea of étendue applies just as much to spectroscopic surveys as to imaging surveys. And the cheap cost of CCD detectors makes possible a huge increase in the field for spectroscopy as well as imaging. Here we describe the basic concept and preliminary design of R. Content, and the science case for an extreme multiplex spectrograph for wide-field 4-m telescopes. The beauty of this instrument design is that there are a number of already existing 4-m telescopes with a 1° field or more where this instrument could be cheaply implemented. With a 1° field, a rate of ~4000 spectra an hour or ~30000 spectra per night is achievable. With a 3 $deg^2$ field, ~12000 spectra per hour or ~100000 spectra per night would be possible. These rates make many surveys possible, especially for cosmology. And the rates that are achievable may be competitive with much more expensive instrument proposals like WFMOS+SKA.

We shall therefore begin by describing the optical design concept of our proposed wide-field spectrographs, and then give some examples of the powerful science that might be addressed by these instruments, before describing our conclusions.

## 2. NG1DF/XMS CONCEPT

The fundamental technology driver here is the cheap cost of CCDs, which now makes it possible to tile a large focal plane cheaply and, coupled with an efficient optical design, still retain high image quality. As noted above, imaging surveys such as Pan-STARRS are already exploiting the étendue of small aperture telescopes with wide fields of view. Here we propose similarly to take advantage of the large étendue of 4-m class Ritchey-Chretien telescopes for extreme multiplex spectroscopy.

*robert.content@durham.ac.uk; phone +44-191-334-3541

### 2.1 Number of slits

The conceptual design of the Next Generation 1° Field (NG1dF) instrument for the AAT 4-m and the Extreme Multiplex Spectrograph (XMS) instrument for the Calar Alto 3.5-m exploits this promising combination of aperture and field of view to allow slits to be placed on ~4000 targets. This estimated number assumes 1.5" width and 10" length; up to 10000 slits could be in principle targeted if the 10" slit width was reduced to a 1.5" aperture, with a few hundred such sky slots to map the sky over the whole mask. The multiplex gain of the instrument would then be even greater. The estimate of 4000 targets also assumes that the spectra would have a length of 200 slit width images on the detector. The width and length of the slits depends on the seeing so a higher spectral resolution and a larger number of sources can be achieved with better seeing.

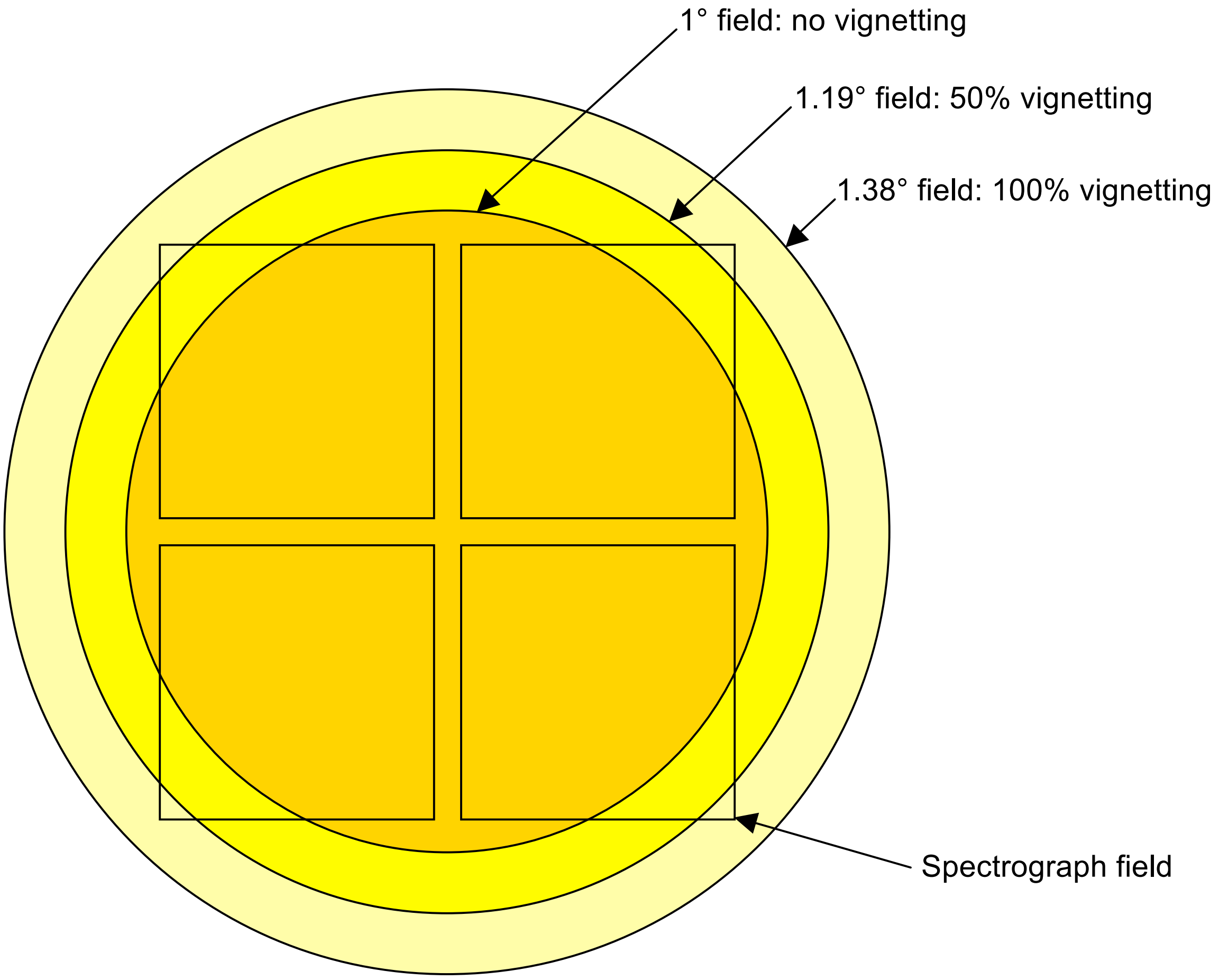

Fig. 1. The field of view of NG1dF for AAT 3.9-m, similar to that to XMS for the Calar Alto 3.5-m. The 1° field is divided into four square subfields.

### 2.2 Spectrograph at prime-focus

To use the large 1° field or more, the spectrograph must be at the prime focus. This put some constrains on its characteristics, mostly putting some limits on its weight and length. A unique spectrograph would easily be too large and too heavy so an array of smaller spectrographs was chosen instead. Our design contains 4 spectrograph units in each of NG1dF and XMS. Each unit has its own detector. A larger number of units would be used at a site of exceptional seeing and/or a telescope with a larger field corrector. Figure 1 shows the present baseline design for the AAT. The subfields are 100 mm wide at the corrector focus which corresponds to 26' on the sky. This size was designed for 2k x 2k detectors. With 4k x 4k detectors, larger subfields may be possible.

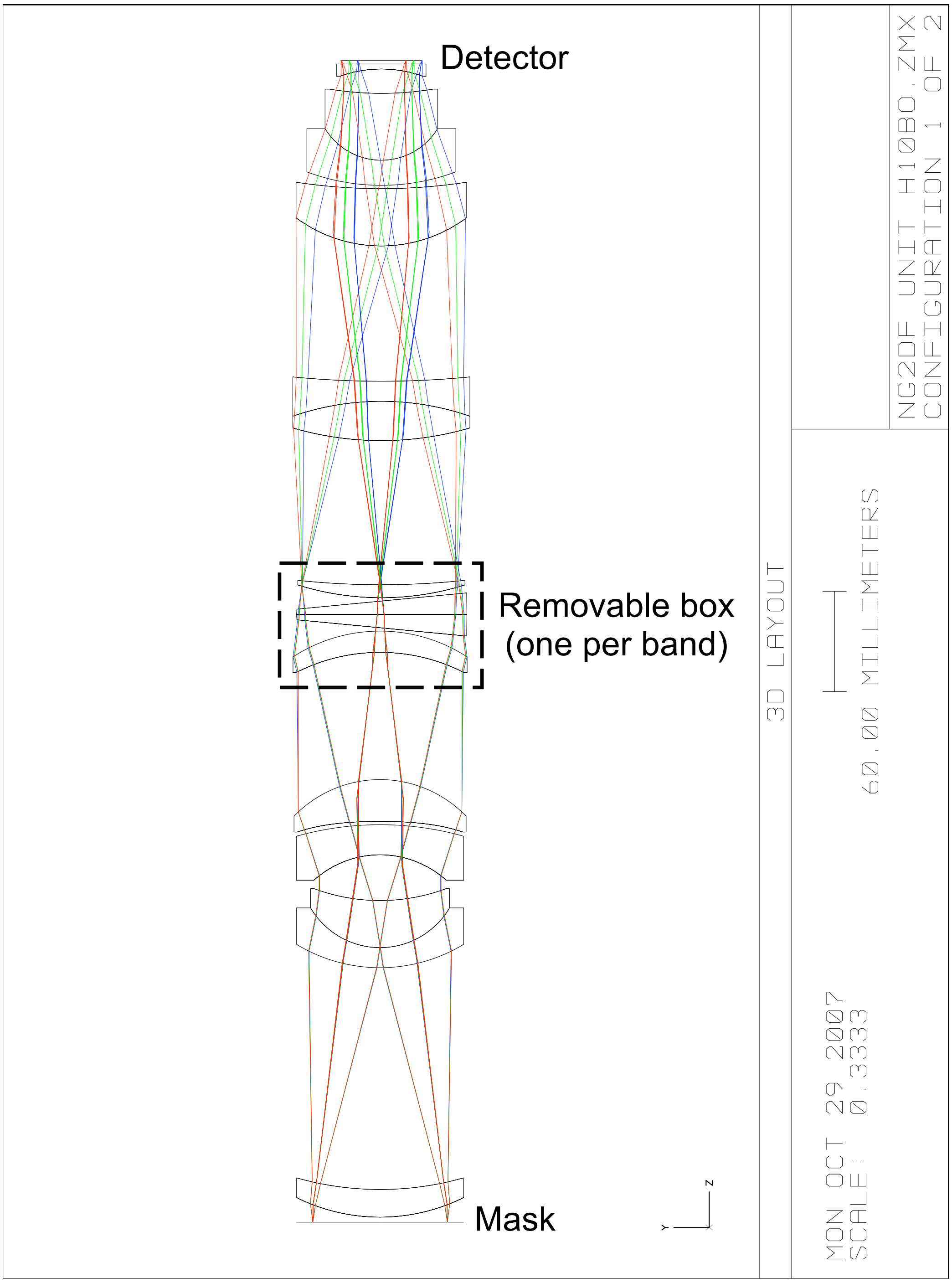

Fig. 2. The design of a single spectrograph for the NG1dF/XMS instrument; four of these are needed to cover the 1° field of classical 4-m telescopes shown in Fig. 1. The box containing the disperser is changed when changing of band or spectral resolution or for imaging.

While there is some vignetting outside the 1° field, there is still a lot of light getting through in the 4 corners of the spectrograph so this region is not lost. With a proper tiling of the sky, some sources in the vignetted area can be observed twice and the light added by software. This would increase the effective surface area of the field of view.

One difficulty with many spectrograph units is to place these units near each other to minimize the gaps in the field. Ideally, all optics in a spectrograph unit would be smaller than the subfield width. This would permit to pack the units tightly. This is achieved in our design by placing a field lens as near as possible to the input focal plane to bend the beams toward the centre (fig. 2). Mechanically, the spectrograph can be made as one structure or each unit can be a separate structure. A combination of both is also possible. While a unique structure would make the spectrograph more rigid for a specific weight, it makes alignments more difficult.

### 2.3 Low cost spectrograph

For this concept to be practical, it is necessary that the price is maintained sufficiently low. A first part of the solution is to avoid the complex mechanism and spectrograph shape that comes with reflective gratings. A transparent disperser combining a transparent grating, as a grism or a VPH, with prisms permits to make the units straight (fig. 2) and makes it possible to place optics very near the disperser which in turn reduces the size of the spectrograph and reduces the aberrations. Different dispersers will have different prism angles and materials depending on the band and spectral resolution. For low spectral resolution, a grism or a grism plus a prism would do. For high resolution, VPH glued to prisms would be the preferred option.

The second part of the solution is to use low cost glasses. This however makes it difficult to design the spectrograph achromatic. To resolve this problem, we take advantage of the need to change the disperser when changing of band. In our solution, one or 2 lenses are changed with the disperser. These lenses can be made small and thin and would fit in a small box with the disperser (fig. 2). A special care must be taken when designing them to avoid the need of tight alignments of the disperser boxes which would increase the cost. One or more of the boxes can be made of lenses only and used for imaging; alternately, the disperser alone can be replaced by some optics in each box.

### 2.4 Changing mechanism

To maintain a low cost, there must be as little mechanisms as possible. The basic concept is to use only one disperser per night and to have it changed by hand during the day. It may however be necessary to have some imaging capability for acquisition. If no practical alternative method can be found, it would be necessary to have a changing mechanism with 2 positions, most probably a slide.

The problem is different for the mask; many of them are necessary every night. The 2 main alternatives are a wheel or a jukebox like mechanism as in GMOS. The wheel would support a smaller number of masks than the jukebox but is simpler mechanically.

### 2.5 Spectral resolution

The basic concept is for a resolution of 10Å between wavelengths of 0.7 μm and 0.9 μm with a 1.5" slit. We could then use 2k x 2k CCDs with 24 μm pixels which gives a sampling of 2 pixels per 1.5" slit image. This sampling does not take advantage of better seeing however. For 0.9" slits, 4k x 4k detectors are necessary. The resolution then becomes 6Å with the same design. Higher resolution are however possible by changing the disperser. The prisms associated with a high resolution disperser would need to have a larger interior angle but the present angle is quite small so it is perfectly possible to at least double the resolution.

Much higher resolutions are also possible but request some significant changes to the design. The camera and collimator could not be maintained in a straight line; the camera would need to be at an angle. This is feasible with a spectrograph that has no more than 6 units but would be problematic with a larger number of units where some are completely encircled by other units. To have a sufficiently large bandwidth at high resolution, it would probably be necessary to have a camera angle that can be changed. This would of course increase the complexity and the cost of the system.

### 2.6 Wavelength range

There is a choice of 2 bands in the present design, each with its own disperser+lenses box. Their wavelength ranges are respectively 0.5 μm to 0.7 μm and 0.7 μm to 0.9 μm. There are then only 2 boxes in the basic design. It is however possible to have other bands. One choice is to have a range of 0.7 μm to 1.0 μm instead of 0.9 μm with a resolution of 15Å instead of 10Å with 1.5" slits. Another would be to have 3 bands between 0.42 μm and 1.0 μm, for example 0.42 μm to 0.55 μm, 0.55 μm to 0.74 μm and 0.74 μm to 1.0 μm.

# 3. SCIENCE CASE

There are many science goals for an instrument of this sort, ranging from galaxy surveys for cosmology at low resolution (R~400) to stellar radial velocity and abundance surveys at 10x higher resolution (R~4000). Here we concentrate on the low resolution cosmological aspects which form our particular interest in the science from these instruments.

### 3.1 Emission and absorption line galaxy redshift surveys at z~0.7.

A prime cosmological goal for the NG1dF/XMS instruments is based on their ability to measure 4000 galaxy redshifts per hour for i<21 absorption-line and i<22 emission-line galaxies at z~0.7 (see Fig. 3). 4000 emission/absorption redshifts an hour means ~30000 redshifts a night or ~6 $x10^6$ galaxy redshifts in a 200 night survey. Such a survey could cover 1000-2000 $deg^2$ of sky and enable powerful new investigations of the clustering of galaxies to be made over a wide range of scales (0.1-1000$h^{-1}$Mpc).

A prime aim would be to measure the scale-length of Baryon Acoustic Oscillations (BAO) as detected in galaxy clustering power spectra and correlation functions. These features are seen as an oscillation in the power spectrum and as a spike in the galaxy correlation function. These features can be used as standard rods and allow tests of cosmological models. In particular, such observations will allow us to probe the equation of state of the vacuum energy, $p=w\rho$. Currently, the spike in the correlation function is tentatively detected in the 2dF Galaxy Redshift Survey of 250000 z~0.1 galaxies [1] and also in the SDSS redshift survey of ~75000 z~0.35 Luminous Red Galaxies [2]. In future bigger galaxy surveys will be needed to measure the BAO scale at higher redshifts and hence track any evolution in the vacuum energy equation of state with redshift. Instruments like NG1dF/XMS therefore will have a crucial role to play in the future of observational cosmology.

There are enough galaxies at the magnitude limits quoted above to fill ~4000 NG1dF/XMS slits since galaxy count data suggest that there are ~4000 galaxies per square degree at i<21 and ~9000 at i<22, 5000 of which will show emission lines. The ~9000 absorption and emission line galaxies available in total will make it more possible to place slits on a subset of ~4000 galaxies. The future for BAO studies is to identify systematics caused by non-linearity in galaxy power spectra that may results in different scale-lengths being measured for different types of galaxy. The high multiplex of NG1dF/XMS will, for example, mean that a choice will no longer have to be made between emission-line galaxies and luminous red galaxies for BAO measurement since both can be observed simultaneously in the same volume and the BAO results compared. Similar comparisons can be made as a function of morphology and luminosity class, given the high numbers of galaxies available.

Other cosmological applications of redshift surveys containing millions of z~0.7 galaxies will include measuring the rate of growth of structure using galaxy clustering redshift-space distortions. Here the flattening of galaxy clustering in redshift-space caused by dynamical infall is used to provide a measure of the infall parameter $\beta=\Omega^{0.6}/b$ and measuring $\beta$ at different redshifts gives the gravitational growth rate. This provides a test of Einstein's gravity independently of geometrical cosmological tests using standard candles and rods, such as BAO. Redshift distortion results can also be used to give an estimate of the masses and hence mass-to-light ratios of galaxy group haloes in CDM models. These allow new tests of the efficiency of the process of galaxy formation as a function of halo mass environment. There are too many examples of other galaxy redshift survey-based projects to detail here. These include the vast array of results that will be available on the topic of the dependence of the galaxy stellar mass and luminosity function on environment and redshift.

### 3.2 Lyman break galaxy redshift surveys at z~3.

More ambitiously, it may be possible for 4-m telescopes to compete effectively on surveying the Universe of galaxies at redshift, z~3. In 4x3hr exposures, spectra for ~4000 r<25 Lyman break galaxies at z~3 (fig. 4) can be measured, producing 100000 z~3 galaxy redshifts in a 50 night survey. Although the exposure time is long, a single mask will produce as many redshifts at z~3 as are currently known from larger telescopes. A prime scientific aim would then again be to measure the BAO scale in the LBG clustering, allowing the first constraints on the dark energy equation of state at z~3. Furthermore, measuring redshift space distortions in the galaxy clustering would give constraints on the gravity model at z=3 and also allow the dependence of the halo M/L ratio on redshift to be estimated and hence the evolution with time of the efficiency of the galaxy formation process. In terms of a survey plan that would include the lower redshift z~0.7 galaxies discussed in Section 3.1, in a total survey time of 250 nights, flexible scheduling might allow the 50 nights with the best seeing to be used to observe fainter LBG targets with 1 arcsec wide slits and the remaining 200 nights could be used for the z~0.7 galaxy redshift survey where 1.5-2 arcsec wide slits might be used. This 250 night survey plan might then realize redshifts for 100000 z~3 LBGs as well as for ~6 million z~0.7 galaxies.

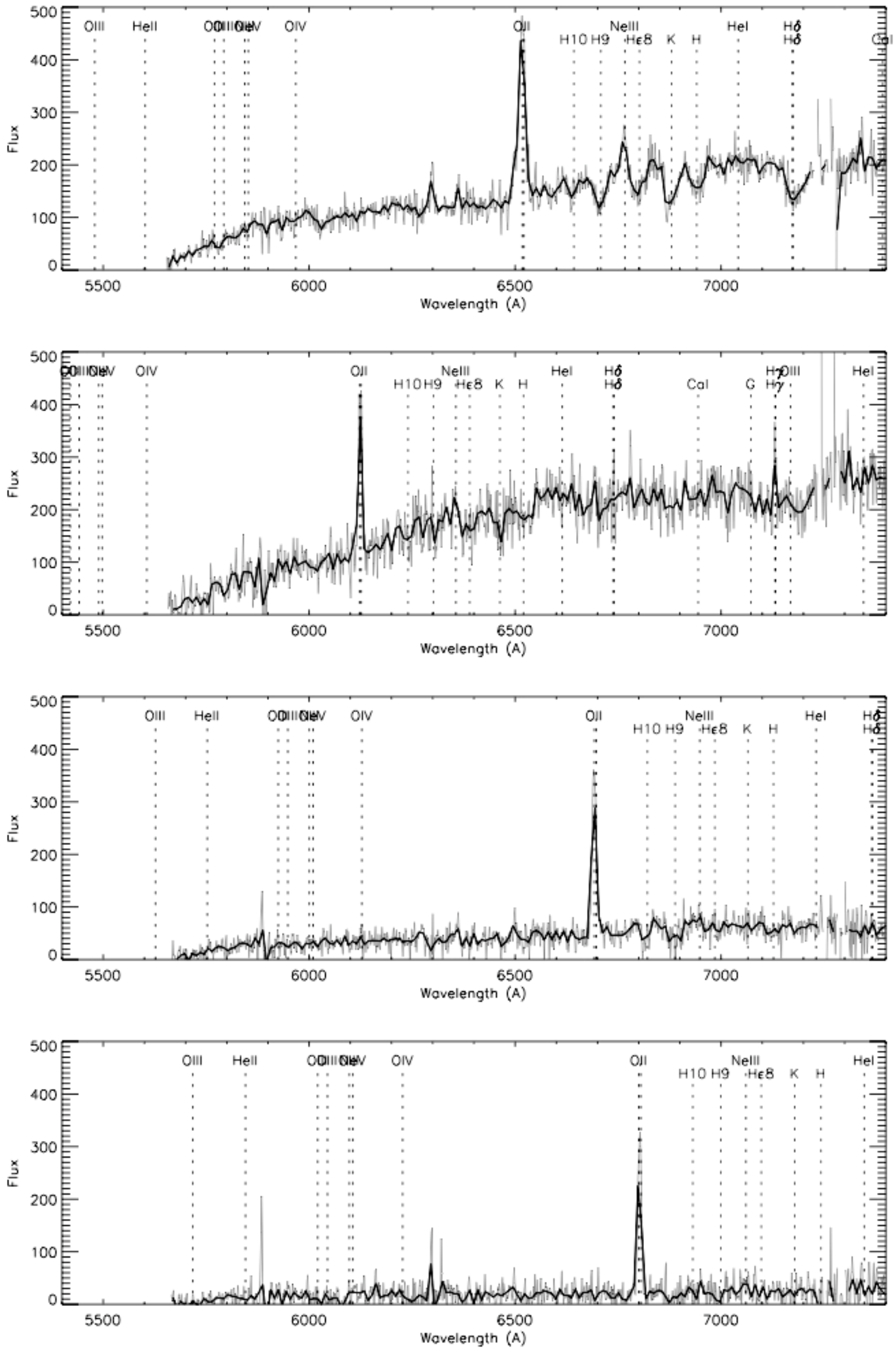

Fig. 3. The exposure time of 1-2 hrs to measure spectra for i<21 galaxies is suggested from exposure times for LRGs with the fibre-fed AAOmega spectrograph at the AAT. The figure shows the galaxy spectra measured by the AAOmega spectrographs at ~5Angstrom resolution and degraded to the ~10Angstrom NG1dF/XMS resolution [3]. These spectra are therefore typical of those that can be achieved with XMS/NG1dF; emission-line and absorption-line features are clearly visible at this resolution.

### 3.3 Calibration of photometric redshifts

A prime use of galaxy surveys like both of those above will be the calibration of photometric redshifts. These are needed to support satellite surveys such as SNAP and also EUCLID which is the new ESA dark energy mission resulting from the merging of SPACE and DUNE [6]. For example, it has been suggested that up to ~100000 spectroscopic redshifts may be needed to calibrate the large photometric redshift surveys planned in the case of the SNAP satellite. Large numbers of galaxies are particularly needed where large galaxy redshift training sets are needed as in the case of Artificial Neural Network (ANN) photometric redshift codes [5]. The NG1dF/XMS absorption and emission line galaxy redshift surveys described above will be uniquely placed to satisfy this huge demand for spectroscopic redshifts for photo-z calibration and would complement the work of these satellites, for example EUCLID would not observe galaxies at a redshift smaller than 0.7.

### 3.4 Stellar velocity and abundance surveys

In its higher resolution mode (R~4000) NG1dF/XMS can also be used for a variety of other purposes. At low Galactic latitudes and in the Galactic bulge, stars brighter than i=15 will have sky densities larger than 4000 per deg$^{-2}$ and can therefore be efficiently observed with this instrument. At these magnitudes, 1-2hr observations will yield an S/N per Angstrom > 50 and thus allow stellar radial velocities and abundances of useful accuracy to be achieved. This will allow Galactic archaeology projects searching for stellar streams in the Galactic bulge and thick disk to proceed. Although the velocity accuracy in will not be high enough in nearby dwarf galaxies, stellar abundance studies will certainly be possible in the higher resolution mode of NG1dF/XMS.

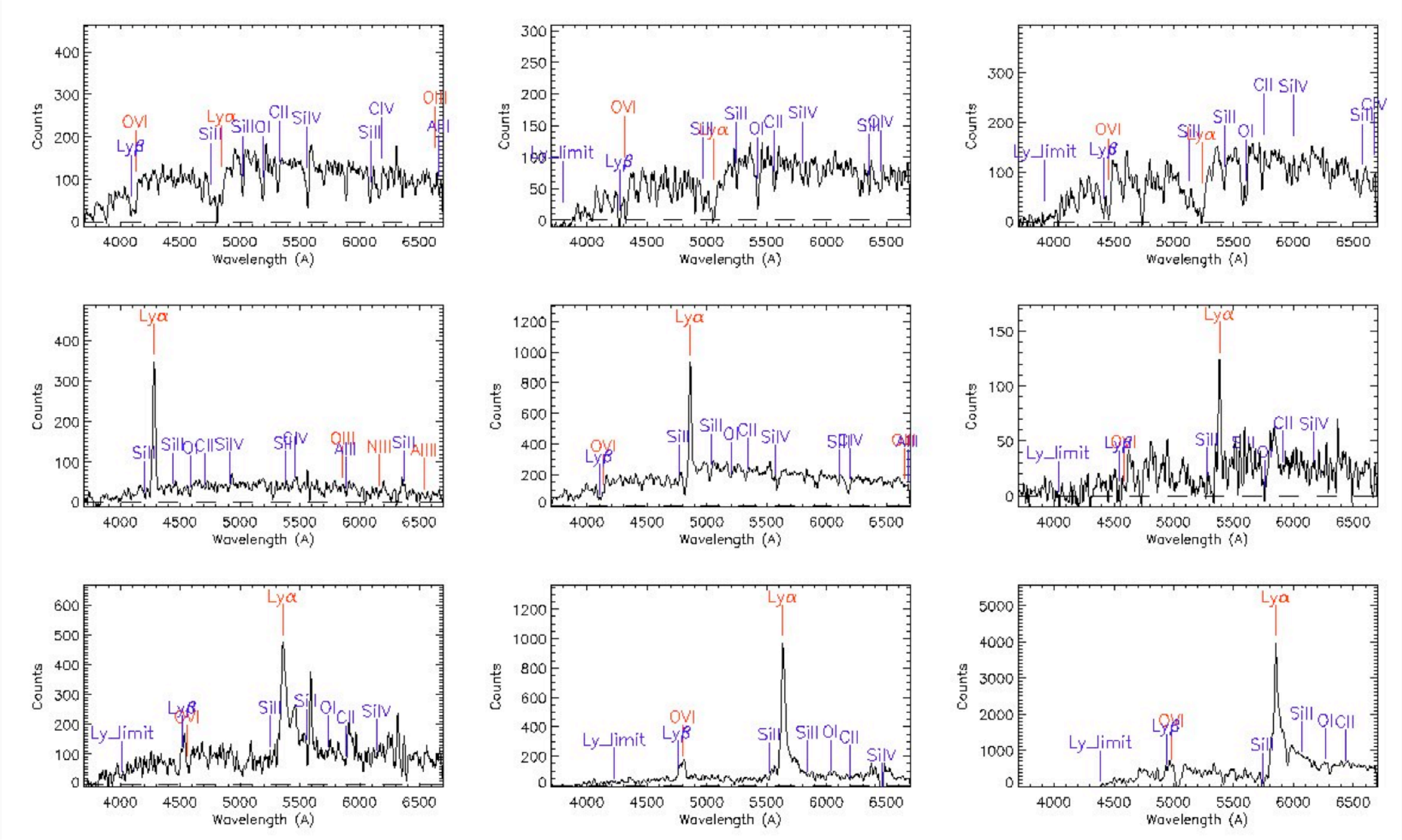

Fig. 4. Seven spectra of Lyman break galaxies at z~3 from a VLT VIMOS survey [4]. Two faint QSO spectra from the same survey are also shown. These spectra taken with the VIMOS LR Blue grism have lower resolution (~20Angstroms) and show that such surveys will be perfectly possible with the default 10Angstrom resolution of the NG1dF/XMS spectrographs.

## 4. CONCLUSIONS

Clearly, spectrographs such as NG1dF/XMS have huge potential impact in terms of the significant science goals they address. There are many ground-based optical and near infra-red imaging surveys coming up such as Pan-STARRS, VST ATLAS+KIDS, ALHAMBRA/PAU, Dark Energy Survey, VISTA Hemisphere Survey and VIKING surveys and also LSST. These will inevitably need spectroscopic follow-up and NG1dF is ideally designed to do this. This spectroscopic follow-up will include the major science programs described above including the measurement of the BAO scale and the measurement of the evolution of the gravitational growth rate. These address major science questions including the nature of dark energy and its evolution and whether Einstein's gravity can explain the observed growth of large-scale structure. However NG1dF/XMS are not "niche instruments" because of the wide variety of science goals they can pursue, of which we have only noted a few examples here.

A major driver is that these instruments can exploit already existing 4-m telescopes with large fields of view and therefore this helps keep costs low. The modular spectrograph design is also aimed at being relatively low-cost as well as maintaining high throughput over a wide field of view. The exact cost is the subject of the ongoing design studies but a preliminary cost estimate (including manpower) is around £2m. The four 4kx4k CCD detectors and associated controllers and Dewars comprise a large fraction (~1/3) of this cost.

We conclude that the future for classically designed 4-m telescopes with large fields may be bright if they are armed with spectrographs such as NG1dF/XMS. Indeed, such telescope-instrument combinations may offer a very cost-effective solution to pursuing the large-scale cosmological surveys that are needed to address the fundamental questions of the dark energy and its evolution with redshift and one which is competitive scientifically with more ambitious projects such as WFMOS and SKA.